\documentclass[conference]{IEEEtran}
\IEEEoverridecommandlockouts

\usepackage[english]{babel}

\usepackage{amsfonts}
\usepackage{circuitikz}
\usepackage{tikz}
\usepackage{standalone} 
\usepackage{tikzscale}
\usetikzlibrary{backgrounds,fit,positioning}
\usetikzlibrary{arrows, patterns,positioning,shapes.geometric,fit,backgrounds}
\usepackage{algorithm} 
\usepackage{algpseudocode} 
\usepackage{mathtools}
\usepackage[thinc]{esdiff}
\usepackage{amsmath,amssymb,color,comment,latexsym} 
\usepackage{bbm} 
\usepackage{textcomp}
\usepackage{pgfplots}
\usepackage{xcolor}
\usepackage{url}
\usepackage{multirow}
\usepackage{graphicx}
\usepackage{caption}
\usepackage{placeins}
\usepgfplotslibrary{units}
\usetikzlibrary{spy,backgrounds}
\usepackage{pgfplotstable}
\usepackage{stmaryrd} 
\usepackage{subcaption}
\usepackage{float}
\usepackage{placeins}  
\usepackage{adjustbox} 
\usepackage{rotating}  

\usepackage{acronym}

\DeclareUnicodeCharacter{200B}{}

\DeclareMathOperator*{\argmax}{argmax}

\DeclareMathOperator{\diag}{diag}

\def\BibTeX{{\rm B\kern-.05em{\sc i\kern-.025em b}\kern-.08em
 T\kern-.1667em\lower.7ex\hbox{E}\kern-.125emX}}

\acrodef{DQN}{Deep $\mathrm{Q}$-Network}
\acrodef{DQL}{Deep $\mathrm{Q}$-Learning}
\acrodef{DL}{Deep Learning}
\acrodef{RL}{Reinforcement Learning}
\acrodef{QL}{$\mathrm{Q}$-Learning}
\acrodef{NN}{Neural Network}
\acrodef{DRL}{Deep Reinforcement Learning}
\acrodef{RIS}{Reconfigurable Intelligent Surfaces}
\acrodef{CSI}{Channel State Information}
\acrodef{mmWave}{Millimeter-wave}
\acrodef{UE}{user}
\acrodef{BS}{base station}
\acrodef{AoD}{angle of departure}
\acrodef{AoA}{angle of arrival}
\acrodef{MDP}{Markov Decision Process}
\acrodef{Eval-Net}{Evaluation Network}
\acrodef{Target-Net}{Target Network}
\acrodef{MSE}{Mean Squared Error}
\acrodef{SNR}{Signal to Noise Ratio}
\acrodef{RSE}{Received Signal Energy}
\acrodef{MIMO}{Multiple-Input Multiple Output}
\acrodef{ULAs}{Uniform Linear Arrays}

\begin{document}

\title{Deep Reinforcement Learning for Blind Passive Beamforming in RIS-Assisted MIMO Systems%
\thanks{Part of this work was funded by the French National Research Agency (ANR-22-PEFT-0005) as part of France 2030 and the NF-YACARI project.}
}


\author{
\IEEEauthorblockN{Oumaïma Bounhar, Thomas Chêne, Ghaya Rekaya-Ben Othman}
\IEEEauthorblockA{\textit{COMELEC, Télécom Paris, IP Paris} \\
Palaiseau, France \\
\{oumaima.bounhar, thomas.chene, ghaya.rekaya\}@telecom-paris.fr}
\and
\IEEEauthorblockN{Oussama Damen}
\IEEEauthorblockA{\textit{ECE depart., University of Waterloo} \\
Waterloo, Canada \\
mdamen@uwaterloo.ca}
}

\maketitle

\begin{abstract}
    Reconfigurable Intelligent Surfaces (RIS) are a promising solution for improving millimeter-wave (mmWave) communications by controlling signal propagation. Yet, selecting the optimal RIS configuration remains challenging due to the high-dimensional search space and signaling overhead. We propose a Deep Reinforcement Learning (DRL) framework for blind passive beamforming, where a Deep $\mathrm{Q}$-Learning (DQL) agent learns an adaptive policy to select near-optimal RIS configurations while sounding only few codewords, without channel state information. Results show that our approach outperforms $\mathrm{Q}$-Learning (QL) and conventional baselines in convergence, performance, and overhead reduction.
\end{abstract}

\label{Introduction}
\section{Introduction}

\IEEEPARstart{M}{illimeter-wave} (mmWave) communications are central to next-generation wireless networks due to their large bandwidth and high spectral efficiency. However, severe path loss and blockages strongly limit communication performance. \ac{RIS} have emerged as a promising solution by enabling dynamic control over the wireless propagation environment through intelligently controlled electromagnetic wave reflections \cite{wuSmartReconfigurableEnvironment2020a}.

\ac{RIS} are either active or passive depending on whether amplification is used \cite{zhangActiveRISVs2023}. Passive \ac{RIS} are attractive for their low cost and energy efficiency \cite{direnzoSmartRadioEnvironments2020}, but configuring many reflecting elements leads to high-dimensional optimization problems. In addition, since passive \ac{RIS} lack transceiver chains, channel estimation is challenging \cite{direnzoSmartRadioEnvironments2020}. Most existing works perform channel estimation by testing a limited number of configurations, by switching some elements on/off over multiple slots \cite{yuan-Zhang-2021, wei-2021}.

\ac{DRL} has been studied for adaptive passive beamforming in \ac{RIS}-aided networks, addressing phase-shift optimization under full, partial, or imperfect \ac{CSI} \cite{marquesguerraRISaidedSystemChannel2022a,huang-2020,saglam-2023}. Separately, \cite{wang2022irs} proposes a Double \ac{DQN}-based approach for \ac{RIS}-assisted multi-user \ac{MIMO} with unblocked direct links and no sub-channel estimation. Blind beamforming, requiring no \ac{CSI} at all, remains comparatively less explored: \cite{blind2025} tunes each \ac{RIS} element individually from a fixed batch of random trials, whereas we adaptively select among a set of pre-designed \ac{RIS} configurations using a learned policy and only a few pilot symbols.

\emph{Contributions:}
In our previous work, we introduced \ac{QL} for our adaptive blind \ac{RIS} configuration protocol without relying on \ac{CSI} \cite{chene-}. However, \ac{QL} does not scale well with the state space and degrades in low- and moderate-\ac{SNR} conditions. In this work, we extend the approach to \ac{DQL}, where \acp{NN} approximate Q-values and generalize over larger continuous state spaces. We focus on passive \ac{RIS} and adopt a blind setting with no channel knowledge. We demonstrate the potential of \ac{DRL} for \ac{RIS} control, optimizing \ac{RIS} phase shift configurations while reducing pilot overhead.
\emph{Notation:} 
We denote $\llbracket1,\mathrm{n}\rrbracket = \{1,2,...,\mathrm{n}\}$, $\mathbf{e}_{\mathrm{i}}$ the $\mathrm{i}$-th canonical basis vector, $(\cdot)^{\mathrm{T}}$ the transpose and $(\cdot)^{\mathrm{H}}$ the Hermitian transpose. Bold lowercase and uppercase letters denote vectors and matrices, respectively, $\mathbf{x}$ and $\mathbf{X}$. We define $\|\mathbf{x}\|_2^2=\mathbf{x}^{\mathrm{H}}\mathbf{x}$, use $\propto$ for proportionality up to normalization, and let $\mathrm{p}(\mathrm{X}\mid \mathrm{Y})$ denote the conditional probability of $\mathrm{X}$ given $\mathrm{Y}$.

\label{Problem Statement}
\section{Problem Statement}
\begin{figure}[h]
\centering
    \includegraphics[width=0.3\textwidth]{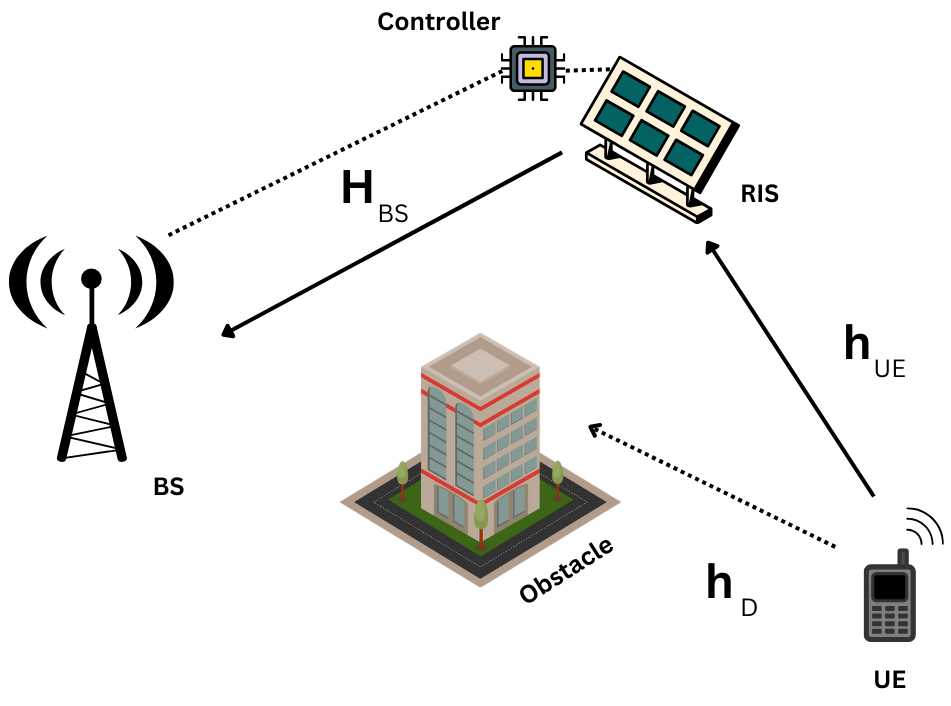}
    \caption{Setup}
    \label{figRIS}
\end{figure}
To provide a first validation of \ac{DQL} for blind passive beamforming in \ac{RIS}-assisted mmWave \ac{MIMO} systems, we intentionally adopt a simplified channel model. More precisely, rather than considering from the outset a fully realistic end-to-end propagation environment, we use a single-path mmWave model as an initial testbed. This choice allows us to isolate the learning behavior of the proposed DRL framework and verify that it can learn efficient \ac{RIS} control policies with limited pilot sounding and without channel knowledge.
We consider a transmission between a single-antenna \ac{UE} and an $\mathrm{M}$-antenna \ac{BS}, assisted by a \ac{RIS} with $\mathrm{N}$ reflecting elements. $\mathbf{h}_{\mathrm{UE}}\in\mathbb{C}^{\mathrm{N}\times 1}$ and $\mathbf{H}_{\mathrm{BS}}\in\mathbb{C}^{\mathrm{M}\times \mathrm{N}}$ denote the respective channels between the \ac{UE} and the \ac{RIS}, and between the \ac{RIS} and the \ac{BS}. We assume that the direct link between the \ac{BS} and the UE, denoted by $\mathbf{h}_{\mathrm{D}}$, is blocked.

We consider a mmWave communication system with half-wave spaced \ac{ULAs}, as adopted in previous works \cite{ningTerahertzMultiUserMassive2021a,xiaoHierarchicalCodebookDesign2016a}. We suppose that each channel has a single path. The channels can be expressed as:
\begin{equation}
	\begin{split}
			\mathbf{H}_{\mathrm{BS}} &=  \mathbf{a}_{\mathrm{M}}(\phi_{1}) \mathbf{a}_{\mathrm{N}}^{\mathrm{H}}(\phi_{2}) \\
    		\mathbf{h}_{\mathrm{UE}} &=  \mathbf{a}_{\mathrm{N}}(\phi_{3}) \mathbf{a}_{1}^{\mathrm{H}}(\phi_{4}) \\
    	\end{split}
\label{eq:channel_equations}
\end{equation}
\noindent where $\mathbf{a}_{\mathrm{N}}(\phi)$ is the steering vector function defined as $\mathbf{a}_{\mathrm{N}}(\phi) = [1,\mathrm{e}^{\mathrm{j}\pi \sin (\phi)},..., \mathrm{e}^{\mathrm{j}\pi (\mathrm{N}-1) \sin (\phi)}]^{\mathrm{T}}$, $\phi_{1}$ and $\phi_{2}$ are the \ac{AoA} and \ac{AoD} of the channel between the \ac{RIS} and the \ac{BS}, $\phi_{3}$ and $\phi_{4}$ are the ones for the channel between the \ac{UE} and the \ac{RIS}. 
\subsection{Received signal}
The signal at the \ac{RIS} will be linearly transformed by a diagonal matrix $\mathbf{\Theta} = \diag(\mathbf{\Phi})$, called reflection matrix, with $ \mathbf{\Phi} =[\mathrm{e}^{\mathrm{j}\theta_0},....,\mathrm{e}^{\mathrm{j}\theta_{\mathrm{N}-1}}]^{\mathrm{T}}$, where the $\theta_{\mathrm{n}}$ are the phase shifts introduced by the \ac{RIS}.
The signal received at the \ac{BS} is:
\begin{equation}
	\mathbf y = \sqrt{\mathrm P}\mathbf{H}_{\mathrm{BS}} \diag(\mathbf\Phi) \mathbf{h}_{\mathrm{UE}} \mathrm{s} + \mathbf w
	\label{eq:signal_1}
\end{equation}
\noindent where $\mathrm P$ is the \ac{UE} transmit power, $\mathrm s\in \mathbb C$ (with $|\mathrm{s}|^2 = 1$) is the signal sent by the \ac{UE}, $\mathbf{w} \sim \mathcal{CN}(0,\sigma_{\mathrm{w}}^2\mathbf{I}_{\mathrm{M}})$ the complex Gaussian additive white noise. 
To communicate, the \ac{RIS} uses the codebook denoted $\mathbf\Phi_{1}^{\mathrm{N_c}} = [\mathbf{\Phi}_1,\ldots,\mathbf{\Phi}_{\mathrm{N_c}}]$, $|\mathbf\Phi_{1}^{\mathrm{N_c}}| = \mathrm{N_c}$, and the objective is to select the optimal \ac{RIS} configuration $\mathbf\Phi_{\text{opt}} \in \mathbf\Phi_{1}^{\mathrm{N_c}}$ to establish high-rate communication. Generally, the system estimates the channel for this communication phase. But in our considered blind protocol, we introduce a parameterization phase: the \ac{UE} sends pilot symbols through the \ac{RIS} using configurations from the pilot codebook $\mathbf{\Psi}_1^{\mathrm{N_p}}=[\mathbf{\Psi}_1,\ldots,\mathbf{\Psi}_{\mathrm{N_p}}]$, with $|\mathbf{\Psi}_1^{\mathrm{N_p}}|=\mathrm{N_p}$. From the received pilot signals and the known pilot configurations, the \ac{BS} infers the posterior probability distribution over the \ac{RIS} configuration codebook elements and determines the optimal \ac{RIS} phase settings, which are then communicated to the \ac{RIS} through the \ac{BS} control link.
\subsection{Optimization Problem}
We define the achievable communication rate $\mathcal{R}$ between the \ac{UE} and the \ac{BS}: 
\begin{equation}
    \mathcal{R} = \log_{2}\!\left( 1 + \frac{\mathrm{P} \, \lVert \mathbf{H}_{\mathrm{BS}} \diag(\mathbf{\Phi}) \mathbf{h}_{\mathrm{UE}} \rVert_{2}^{2}}{\sigma_{\mathrm{w}}^{2}} \right)
\label{eq:rate}
\end{equation}
Our main objective is to maximize the rate $\mathcal{R}$, which leads to maximizing the norm of the combined channel $\mathbf{h}(\mathbf{\Phi})=\mathbf{H}_{\mathrm{BS}}\diag(\mathbf{\Phi})\mathbf{h}_{\mathrm{UE}}$ over the codebook of communication  $\mathbf\Phi_{1}^{\mathrm{N_c}}$: 
\begin{equation}
    \mathbf{\Phi}_{\text{opt}} = \argmax_{\mathbf\Phi \in \mathbf{\Phi}_{1}^{\mathrm{N_c}}} \left( \lVert \mathbf{h}(\mathbf\Phi) \rVert_{2}^{2} \right).
\label{eq:optim_pb}
\end{equation}

\section{Proposed Approach}
For simplicity, we will write $\mathbf H = \{\mathbf{H}_{\mathrm{BS}},\mathbf{h}_{\mathrm{UE}}\}$ to represent the overall unknown channel.
\subsection{Classification}
For a channel realization $\mathbf H$, let $\mathbf\Phi(\mathbf H)$ denote the optimal communication codeword in $\mathbf\Phi_{1}^{\mathrm{N_c}}$:
\begin{equation}
	\mathbf\Phi(\mathbf H) = \{\mathbf\Phi \in \mathbf\Phi_{1}^{\mathrm{N_c}} : \mathbf\Phi = \argmax_{\mathbf\Phi \in \mathbf\Phi_{1}^{\mathrm{N_c}}} \bigl(\lVert\mathbf{h}(\mathbf\Phi) \rVert_2^2\bigr)\}.
\end{equation}
We introduce a probability vector associated with the communication codebook. Each entry corresponds to the probability that a given communication codeword $\mathbf\Phi_{\mathrm{i}}$, $ \mathrm{i} \in \llbracket 1,\mathrm{N_c} \rrbracket$ is the optimal one for the observed channel realization:
\begin{equation}
    \mathbf{p}(\mathbf\Phi(\mathbf{H})) = \Big[\mathbf{p}(\mathbf\Phi(\mathbf{H}) = \mathbf\Phi_{1}), \ldots, \mathbf{p}(\mathbf\Phi(\mathbf{H}) = \mathbf\Phi_{\mathrm{N_c}})\Big].
 \label{eq:probability}
\end{equation}
Our optimization problem~(\ref{eq:optim_pb}) results in a classification problem where we are looking for the policy that assigns the best communication codeword to a channel realisation.
\subsection{Protocol}



Since no prior information about the channel is available, we start with a uniform probability distribution over the communication codebook $\mathbf{\Phi_{1}^{\mathrm{N_c}}}$. At each iteration, the system collects a measure of the RIS configuration, updates its belief about the optimal codeword $\mathbf\Phi_{\text{opt}} \in \mathbf{\Phi_{1}^{\mathrm{N_c}}}$, and stops once a sufficient level of precision is reached.
\begin{itemize}
    \item \textbf{Initialization:}  
    Set uniform prior distribution over the communication codebook $\mathbf{\Phi_{1}^{\mathrm{N_c}}}$:  
    \begin{equation}
        \mathbf{p}(\mathbf\Phi(\mathbf{H})) = \Big[\tfrac{1}{\mathrm{N_c}}, \tfrac{1}{\mathrm{N_c}}, \ldots, \tfrac{1}{\mathrm{N_c}}\Big].
    \label{eq:p_init}
    \end{equation}
    \item \textbf{Iteration $\mathrm{k} = 1, \dots, \mathrm{L_p}$:}
    \begin{itemize}
        \item The \ac{UE} sends a pilot symbol $\mathrm{s_k}$.  
        \item The \ac{RIS} uses the index of a pilot codeword $\pi_{\mathrm{k}} \in \llbracket 1,\mathrm{N_p} \rrbracket$, i.e., $\mathbf\Psi_{\pi_{\mathrm{k}}} \in \mathbf{\Psi}_{1}^{\mathrm{N_p}}$ defined by the \ac{BS}. 
        \item The \ac{BS} receives the signal:  
        \begin{equation}
            \mathbf y_{\mathrm{k}} = \sqrt{\mathrm{P}}\mathbf{H}_{\mathrm{BS}}\diag(\mathbf\Psi_{\pi_{\mathrm{k}}})\mathbf{h}_{\mathrm{UE}}\mathrm{s_k} + \mathbf{w_{\mathrm{k}}}.
        \end{equation}
        \item The \ac{BS} computes the \ac{RSE}:  
        \begin{equation}
            \mathrm Y_{\mathrm{k}} = \mathbf y_{\mathrm{k}}^{\mathrm{H}} \mathbf y_{\mathrm{k}} = \lVert \mathbf y_{\mathrm{k}} \rVert_2^2.
        \label{eq:RSE}
        \end{equation}
        \item Update the posterior probability distribution as described in Section~(\ref{approx_proba}):  
        \begin{equation}
            \mathbf{p}(\mathbf\Phi(\mathbf H) \mid [\mathbf Y_{1}, \ldots,  \mathbf Y_{\mathrm{k}}], [ \mathbf{\Psi}_{\pi_1}, \ldots,  \mathbf{\Psi}_{\pi_{\mathrm{k}}}]).
        \label{eq:update_proba}
        \end{equation}
        \item Based on the updated posterior, the \ac{BS} selects the index $\pi_{\mathrm{k+1}} \in \llbracket 1,\mathrm{N_p} \rrbracket$ that will be used as the next pilot codeword by the \ac{RIS}.
    \end{itemize}
    \item \textbf{Stopping criteria:}  
    We fix a precision level $1-\delta$, with $\delta \in [0,1]$ and a maximum pilot budget $\mathrm{L^{*}}$.
    If:
    \begin{equation}
        \exists \mathrm{i} \in \llbracket 1,\mathrm{N_c} \rrbracket \mid \mathrm{p}(\mathbf\Phi(\mathbf H) = \mathbf\Phi_{\mathrm{i}} \mid \mathbf Y_{1}^{\mathrm{k}}, \mathbf\Psi_{\pi_1}^{\pi_{\mathrm{k}}}) > 1-\delta
    \label{eq:precision}
    \end{equation}
    then declare $\mathbf\Phi_{\text{dec}} = \mathbf\Phi_{\mathrm{i}}$ as the optimal codeword $\mathbf\Phi_{\text{opt}}$.
\end{itemize}
We not only want to optimize the transmission rate~(\ref{eq:optim_pb}) but also reach the stopping criteria~(\ref{eq:precision}) as quickly as possible to reduce the signaling overhead.
\subsection{Markov Decision Process}
\label{MDP}
Since the \ac{RIS} codeword selection problem is sequential: each pilot transmission allows the system to update its belief regarding the optimal codeword and adaptively select the subsequent configuration, we formulate our optimization problem~(\ref{eq:optim_pb}) as an \ac{MDP}. By defining the state space $\mathbf S$ over the belief states (probability distributions), the \ac{MDP} framework effectively captures the dynamic evolution of the system through its action space $\mathbf A$, transition probability $\mathbf P$, and rewards $\mathbf R$~\cite{sutton2018reinforcement}:
\begin{itemize}
    \item \textbf{State Space}: $\mathrm{S} \in \mathbf S$ denotes a state from the state space. Each state $\mathrm{S}$ represents the knowledge acquired after testing $\mathrm{L_p}$ \ac{RIS} configurations $\mathbf{\Psi}_{\pi_1}^{\pi_{\mathrm{L_p}}}$ and collecting measurements $\mathrm{Y}_{1}^{\mathrm{L_p}}$. We write a state $\mathrm{S} = \mathbf p(\mathbf\Phi(\mathbf H)| \mathrm{Y}_{1}^{\mathrm{L_p}},\mathbf{\Psi}_{\pi_1}^{\pi_{\mathrm{L_p}}})$, i.e., the posterior probability vector associated with each communication codeword to be $\mathbf\Phi_{\text{dec}}$. A state is a \textbf{terminal state} ($\mathbbm{1}_{\mathrm{T}}=1$) when~(\ref{eq:precision}) is met; exhausting the pilot budget $\mathrm{L}^{*}$ first only \textbf{truncates} the search ($\mathbbm{1}_{\mathrm{T}}=0$).

    \item \textbf{Action Space}: $\mathrm{A} \in \mathbf{A}$ is an action from the action space. The action space is defined by $\llbracket 1,\mathrm{N_p} \rrbracket$, the indices of the codebook of pilots $\mathbf{\Psi}_{1}^{\mathrm{N_p}}$. Each action corresponds to selecting the index of a specified codeword $\mathbf{\Psi}_{\pi_{\mathrm{k}}}$, $\pi_{\mathrm{k}} \in \llbracket 1,\mathrm{N_p} \rrbracket$, which will be the next configuration to test.

    \item \textbf{Reward}: $\mathrm{R}(\mathrm{S},\mathrm{S}',\mathrm{A}) \in \mathbb{R}$ is the reward received after the transition from state $\mathrm{S}$ to state $\mathrm{S}'$ under action $\mathrm{A}$. It equals $0$ at a terminal state and $-1$ otherwise. We write $\mathrm{R_{t+1}} \triangleq \mathrm{R}(\mathrm{S_t},\mathrm{S_{t+1}},\mathrm{A_t})$ for the reward received at step $\mathrm{t}$.

    \item \textbf{Transition probability}: $\mathbf{P}(\mathrm{S},\mathrm{S}', \mathrm{A}) = \mathbf{P}(\mathrm{S}_{\mathrm{t}+1} = \mathrm{S}' \mid \mathrm{S_t} = \mathrm{S}, \mathrm{A_t} = \mathrm{A})$ is the probability to move from state \(\mathrm{S}\) at time \(\mathrm{t}\) to state \(\mathrm{S}'\) at time \(\mathrm{t}+1\) after taking action \(\mathrm{A}\).
\end{itemize}
\subsection{Optimal Policy}
The goal is to reach a terminal state where the probability of correct classification is within a fixed degree of precision $\delta \in [0;1]$, as quickly as possible. Let $\mathrm{L}_{\min}$ denote the required number of iterations. Since every pilot receives $\mathrm{R_{t+1}}=-1$ except the last one, which reaches the terminal state, receives $0$ and ends the episode (the terminal state is absorbing, so all later rewards are $0$ and the sum below is finite),
\begin{equation}
    \mathrm{L}_{\text{min}} = 1-\sum_{\mathrm{t}=0}^{\infty} \mathrm{R_{t+1}}.
\end{equation}
Therefore, minimizing the path length is equivalent to maximizing the cumulative reward. The objective is thus to find the optimal policy $\pi$ maximizing the expected return under channel and noise randomness. We rewrite our objective:
\begin{equation}
    \pi^{*} = \arg\max_{\pi} \mathbb{E}_{\mathbf{H},\mathbf{w}}\left[\sum_{\mathrm{t}=0}^{\infty} \mathrm{R_{t+1}} \right]
\label{eq:L_min_*}
\end{equation}
\section{Reinforcement Learning Formulation}
\subsection{Q-Learning Algorithm}
\label{sec:ql}
The \ac{QL} algorithm is a \ac{RL} method that learns the optimal policy by iteratively updating a Q-table, which stores \textbf{Q-values} for each state-action pair. It was a natural first choice in our earlier study as the action space is discrete and tabular \ac{QL} provides an interpretable baseline for finite-state decision problems~\cite{sutton2018reinforcement}. The \textbf{Q-function} represents the expected cumulative reward an agent can achieve from a given state $\mathrm{S_t}$​, taking action $\mathrm{A_t}$​, and following the optimal policy ever after:
\begin{equation}
    \mathrm{Q}(\mathrm{S_t},\mathrm{A_t})=\mathbb{E}_{\mathbf H,\mathbf w}[\textstyle\sum_{\mathrm{k}=0}^{\infty}\gamma^{\mathrm{k}} \mathrm{R}_{\mathrm{t}+\mathrm{k}+1}\mid \mathrm{S_t}=\mathrm{S},\ \mathrm{A_t}=\mathrm{A}]
\label{eq:q_fct}
\end{equation} with $\gamma \in [0,1]$, the discount factor, controls the weight of future rewards compared to immediate ones.
To maximize the Q-function, \ac{QL} follows a recursive update rule based on the Bellman equation~(\ref{eq:bellman}), which ensures that the Q-values are updated by incorporating rewards received from future states. The update rule is defined as:
\begin{equation}
    \mathrm{Q}(\mathrm{S_t}, \mathrm{A_t}) \leftarrow (1-\alpha)\,\mathrm{Q}(\mathrm{S_t}, \mathrm{A_t}) + \alpha\!\left[{\mathrm{R}}_{\mathrm{t}+1} + \gamma \max_{\mathrm{A}} \mathrm{Q}(\mathrm{S}_{\mathrm{t}+1}, \mathrm{A}) \right]
\label{eq:bellman}
\end{equation}
with $\alpha \in [0,1]$, the learning rate, controls the weight given to new information relative to past Q-values. During training, the agent explores $(\mathbf S,\mathbf A,\mathbf R,\mathbf P)$ and updates Q-values to reach the terminal state in fewer steps. In our previous work~\cite{chene-}, the state space was discretized and transitions were determined by the Euclidean distance to predefined states, an approach that becomes unreliable in high dimensions due to possible misclassification of similar codewords. We therefore replace the Q-table with a \ac{DQN} that generalizes over continuous state spaces, following the standard use of neural function approximation in large-scale reinforcement learning~\cite{mnih2015human}.
\vspace{-1.1mm}
\subsection{Deep Q-Learning Algorithm}
\ac{DQL} uses two \ac{NN}: The first one is the \ac{Eval-Net} parameterized by the \ac{NN} weights $\theta^{(\mathrm{eval})}$ and the second one is the \ac{Target-Net} parameterized by \ac{NN} weights $\theta^{(\mathrm{target})}$. The Q-function approximation can thus be expressed as:
\begin{equation}
\mathrm{Q}_{\text{eval}}(\mathrm{S}, \mathrm{A}; \theta^{(\mathrm{eval})}) \approx \mathbb{E}\left[\mathrm{R} + \gamma \max_{\mathrm{A}'} \mathrm{Q}_{\text{target}}(\mathrm{S}', \mathrm{A}'; \theta^{(\mathrm{target})})\right],
\end{equation}
The training process of the \ac{DQN} consists of:
\begin{itemize}
    \item \textbf{Epsilon-Greedy Exploration:} The agent selects actions based on an epsilon-greedy policy:
    {\small
    \begin{equation}
        \Psi =
        \begin{cases}
            \argmax\limits_{\Psi} \mathrm{Q}_{\text{eval}}(\mathrm{S}, \Psi; \theta^{(\text{eval})}) 
            & \text{with probability } (1 - \epsilon), \\[6pt]
            \text{random action from } \mathbf{\Psi} 
            & \text{with probability } \epsilon.
        \end{cases}
        \label{eq:greedy_policy}
    \end{equation}
    }
    \item \textbf{Experience Replay:} A replay buffer \(\mathcal{M}\) stores transitions \((\mathrm{S}, \mathrm{A}, \mathrm{R}, \mathrm{S}', \mathbbm{1}_{\mathrm{T}})\), and random mini-batches from \(\mathcal{M}\) are sampled during training to decorrelate experiences.
    \item \textbf{Target Network Update:} To improve training stability, the \ac{Target-Net} is updated using a soft-update rule:
    \begin{equation}
        \theta^{(\text{target})} \leftarrow \tau\,\theta^{(\text{eval})} + (1-\tau)\,\theta^{(\text{target})},
    \label{eq:soft_update}
    \end{equation}
    where $\tau \ll 1$ is a smoothing factor that controls the rate of updates from the evaluation network to the target network.
\end{itemize}
\subsection{Loss Function and Network Optimization}
The Q-network is trained by minimizing the \ac{MSE} loss, defined as:
\begin{equation}
    \mathrm{L}(\theta)=\mathbb{E}\!\left[\left(\mathrm{q}_{\text{target}}-\mathrm{Q}_{\text{eval}}(\mathrm{S},\mathrm{A};\theta^{(\text{eval})})\right)^2\right],
\label{eq:loss}
\end{equation}
\noindent with:
\begin{equation}
\mathrm{q}_{\text{target}}=\mathrm{R}+\gamma\,(1-\mathbbm{1}_{\mathrm{T}}) \max_{\Psi} \mathrm{Q}_{\text{target}}(\mathrm{S}',\Psi;\theta^{(\text{target})}).
\end{equation}
$\mathrm{Q}_{\text{eval}}$ predicts current Q-values, while $\mathrm{Q}_{\text{target}}$ provides stable targets that prevent divergence.
\begin{algorithm}[!b]
    \begin{algorithmic}[1]
        \State \textbf{Input:} Dataset $\mathcal D$, Action space $\mathbf\Psi_1^{\mathrm{N_p}}$, Reward function $\mathbf{R}$, Precision $\delta$, Target update factor $\tau$
        \State Initialize $\mathrm{Q}_{\text{target}}$, $\mathrm{Q}_{\text{eval}}$, replay memory $\mathcal{M}$
        \For{each episode}
            \For{each channel realization}
                \State Sample channel realization $\mathbf{H}_{\ell}$ from $\mathcal D$
                \For{each time step per channel}
                    \State Initialize state $\mathrm{S} = \mathrm{S}_{\text{init}}$
                    \For{{$\mathrm{L_p} = 1,\ldots,\mathrm{L^{*}}$}}
                        \State Select pilot codeword index $\pi_{\mathrm{L_p}}$~(\ref{eq:greedy_policy})
                        \State Observe the measure $\mathrm Y_{\mathrm{L_p}}$~(\ref{eq:RSE})
                        \State Update the posterior probability with~(\ref{eq:approx_proba}):
                        \begin{equation}
                            \mathrm{S}' = \mathbf p(\mathbf\Phi \mid \mathbf Y_1^{\mathrm{L_p}}, \mathbf\Psi_{\pi_1}^{\pi_{\mathrm{L_p}}})
                        \end{equation}
                        \State Compute $\mathrm{R}(\mathrm{S}, \mathrm{S'},\mathbf\Psi_{\pi_{\mathrm{L_p}}})$ and flag $\mathbbm{1}_{\mathrm{T}}$
                        \State Store $(\mathrm{S}, \mathbf\Psi_{\pi_{\mathrm{L_p}}}, \mathrm{R}, \mathrm{S}',\mathbbm{1}_{\mathrm{T}})$ in $\mathcal{M}$
                        \State $\mathrm{S} \leftarrow \mathrm{S}'$; \textbf{stop} if $\mathbbm{1}_{\mathrm{T}}=1$
                    \EndFor
                    \If{$\mathcal{M}$ is full}
                        \State Sample mini-batch from $\mathcal{M}$, 
                        \State Update $\mathrm{Q}_{\text{eval}}$ by computing the loss~(\ref{eq:loss})
                        \State Soft-update the \ac{Target-Net} using~(\ref{eq:soft_update})
                    \EndIf
                \EndFor
            \EndFor
        \EndFor
    \end{algorithmic}
\caption{Deep Q-Learning for Optimal Codebook Selection}
\label{alg:Alg3}
\end{algorithm}
To find the optimal policy, we train an agent over the environment described in Section~(\ref{MDP}), following Algorithm~\ref{alg:Alg3}. The agent always starts at the initial state $\mathrm{S}_{\text{init}} = \mathbf{p}(\mathbf\Phi(\mathbf{H}))$ as defined in~(\ref{eq:p_init}), where we assumed the probability is uniformly distributed among the communication codebook $\mathbf\Phi_{1}^{\mathrm{N_c}}$, and aims to reach a terminal state within the limited pilot budget $\mathrm{L^{*}}$.
\subsection{Deep Q-Network Inference Complexity}
\label{sec:complexity}
The proposed \ac{DQN} approximates the $\mathrm{Q}$-function with a fully connected \ac{NN} of $\mathrm{h}$ hidden layers of $\mathrm{d}$ neurons. For an $\mathrm{n}_{\text{in}}$-dimensional state $\mathrm{S}$ and $\mathrm{n}_{\text{out}}$ actions, a forward pass costs
\begin{equation}
    \mathcal{O}\!\big(\mathrm{n}_{\text{in}}\mathrm{d} + (\mathrm{h}-1)\mathrm{d}^2 + \mathrm{d}\,\mathrm{n}_{\text{out}}\big),
\end{equation}
which reduces to $\mathcal{O}(\mathrm{d}^2)$ for the configuration of Section~\ref{sec:dql_params} ($\mathrm{d}=256$, $\mathrm{h}=2$, $\mathrm{n}_{\text{in}}, \mathrm{n}_{\text{out}} \ll \mathrm{d}$). Adding the posterior update of Section~\ref{approx_proba}, computed in $\mathcal O(|\mathcal D|\mathrm{L_p})$ per step, the total online cost is $\mathcal{O}(\mathrm{d}^2 + |\mathcal D|\mathrm{L_p})$ per decision, with only marginal dependence on the codebook sizes. Training is performed entirely \textit{offline} over varying channel realizations; online, only a forward pass of the network is required to determine the optimal action.
\label{sec:Numerical Results}
\section{Numerical Results}
In this section, we compare the numerical performance of the proposed \ac{DQL} model with \ac{QL} and state-of-the-art benchmarks.
Simulations consider a \ac{BS} with $\mathrm{M}=64$ antennas, a \ac{UE} with a single-antenna, and a \ac{RIS} with $\mathrm{N}=100$ reflective elements. The \ac{AoA} and \ac{AoD} of all links are drawn independently from $\mathcal{U}[0,2\pi]$, and the hierarchical codebook follows classical beamforming designs~\cite{xiaoHierarchicalCodebookDesign2016a}. The first training of the model employs $\mathbf{\Psi}_{1}^{\mathrm{N_p}}=14$ pilot beams (wide and narrow) and $\mathbf{\Phi}_{1}^{\mathrm{N_c}}=8$ communication beams. To evaluate scalability, a larger codebook with $\mathbf{\Psi}_{1}^{\mathrm{N_p}}=30$ and $\mathbf{\Phi}_{1}^{\mathrm{N_c}}=16$ beams is also considered. This comparison highlights the model’s ability to maintain convergence stability as the codebook size increases. 

    
\subsection{Dataset Generation}
To train the \ac{DQL} model, we generate a dataset $\mathcal{D}$ containing pilot-based observations and their corresponding terminal-state labels over multiple channel realizations and \ac{SNR} values (0, 5, 10, 20 dB).
\begin{itemize}
    \item \textbf{Channel Model:} Each channel realization $\mathbf H_{\ell}$ is randomly generated following equations defined in~(\ref{eq:channel_equations}) and repeating this process over many realizations (Monte Carlo sampling) for each \ac{SNR} value to build the dataset.
    \item \textbf{Received measures from RIS configurations:}  
    \begin{equation}
        \mathbf X = [\mathbf X(\mathbf H_1),\ldots,\mathbf X(\mathbf H_{\mathrm{N}_{\text{Dataset}}})],
    \label{eq:Xmeasures_dataset}
    \end{equation}
    where $\mathbf X(\mathbf H_{\ell})$ is the vector of received measures from the \ac{RIS} configurations under channel realization $\mathbf H_{\ell}$. For each pilot codeword $\mathbf\Psi_{\mathrm{j}}, \mathrm{j} \in \llbracket 1,\mathrm{N_p} \rrbracket$ from the pilots codebook $\mathbf\Psi_{1}^{\mathrm{N_p}}$, the noisy \ac{RSE} is computed as in equation~(\ref{eq:RSE}) then stored in the dataset, forming the actual observations.
    \item \textbf{Real terminal states:}  
    The optimal communication codeword $\mathrm{c}^*$ is selected from the communication codebook $\mathbf\Phi_{1}^{\mathrm{N_c}}$ (narrow beams) using the noiseless received power:
    \begin{equation}
        \mathrm{c}^{*} = \argmax_{\mathrm{i} \in \llbracket 1,\mathrm{N_c} \rrbracket} \|\mathbf h(\mathbf\Phi_{\mathrm{i}})\|_2^2,
    \end{equation}
    leading to the terminal state represented as a canonical basis vector:
    \begin{IEEEeqnarray}{rCl}
        \mathbf{S}_{\mathrm{T}}(\mathbf H_{\ell})
        &=& \mathbf{e}_{\mathrm{c}^*}, 
    \end{IEEEeqnarray}
        where $\mathbf{e}_{\mathrm{c}^*}$ is a vector of all zeros except for a $1$ at the index corresponding to the optimal beam $\mathrm{c}^*$.
    \label{eq:real_terminal_state}
    Collecting these for all channels gives:
    \begin{equation}
        \mathbf S_{\mathrm{T}_{\text{Real}}} = [\mathbf S_{\mathrm{T}}(\mathbf H_1),\ldots,\mathbf S_{\mathrm{T}}(\mathbf H_{\mathrm{N}_{\text{Dataset}}})].
    \label{eq:all_real_terminal_states}
    \end{equation}
    Let $\mathrm{N}_{\min}$ denote the minimal number of dataset realizations per \ac{SNR} value and per communication codeword. In total, $\mathrm{N}_{\text{Dataset}} = |\mathcal D| = 2\mathrm{N}_{\min}\mathrm{N_c}$ realizations are drawn, equally split among the \ac{SNR} values.
    \item \textbf{Final dataset structure:}  
    Each channel realization thus contributes a pair $(\mathbf X(\mathbf H_{\ell}),\mathbf S_{\mathrm{T}}(\mathbf H_{\ell}))$, where $\mathbf X(\mathbf H_{\ell})$ contains the noisy pilot-based observations and $\mathbf S_{\mathrm{T}}(\mathbf H_{\ell})$ encodes the true terminal state (optimal communication codeword). The final dataset is organized as:
    \[
        \mathcal{D} = (\mathbf X,\mathbf S_{\mathrm{T_{Real}}}),
    \]
    which jointly provides the training samples and labels for the learning algorithms.
\end{itemize}
\subsection{Approximation of the Posteriori Probability}
\label{approx_proba}
To move from a state \(\mathrm{S}\) at time $\mathrm{t}$ to a state \(\mathrm{S}'\) at time $\mathrm{t}+1$, we update the posterior probability using the Euclidean distance between the samples $\mathbf Y_{1}^{\mathrm{L_p}}$ received after $\mathrm{L_p}$ iterations since the last prior update and the corresponding observations stored in the dataset $\mathcal{D}$:
\begin{IEEEeqnarray}{rCl}
    &&\mathrm{p}(\mathbf\Phi(\mathbf H) = \mathbf\Phi_{\mathrm{i}} \mid \mathbf Y_{1}^{\mathrm{L_p}},\mathbf{\Psi}_{\pi_1}^{\pi_{\mathrm{L_p}}}) \propto \nonumber\\
    &&\mathrm{p}(\mathbf\Phi(\mathbf H) = \mathbf\Phi_{\mathrm{i}} \mid \mathbf Y_{1}^{\mathrm{L_p}-1},\mathbf{\Psi}_{\pi_1}^{\pi_{\mathrm{L_p}-1}}) \nonumber \\
    &&\cdot \sum_{\mathbf H_{\ell} \, / \, \mathbf S_{\mathrm{T}}(\mathbf H_{\ell})[\mathrm{i}] = 1}
    \frac{1}{\left\lvert \mathrm X(\mathbf H_{\ell})[\pi_{\mathrm{L_p}}] - \mathrm Y_{\mathrm{L_p}} \right\rvert^2}
    \label{eq:approx_proba}
\end{IEEEeqnarray}
Here, \(\mathbf{\Psi}_{\pi_1}^{\pi_{\mathrm{L_p}}}\) denotes the sequence of decided pilot codewords up to time \(\mathrm{L_p}\). Since the expression is proportional (\(\propto\)), the probability is normalized so that its components sum to one. Note that this Euclidean distance is computed between received observations against stored \acp{RSE}, not between the posterior and a set of predefined states, which is the quantization discussed in Section~\ref{sec:ql} that \ac{DQL} avoids.  Although under the single-path model~(\ref{eq:channel_equations}) the optimal configuration depends only on the angular difference between the \ac{UE}-\ac{RIS} and \ac{RIS}-\ac{BS} paths 
($\sin\phi_3 - \sin\phi_2$),
the agent is blind to this angle and operates on the posterior, a probability vector of size $\mathrm{N_c}$ that grows with the codebook.
\subsection{Deep Q-Learning parameters}
\label{sec:dql_params}
The \ac{DQL} model uses two hidden layers of 256 neurons with ReLU and LayerNorm to map the posterior probability state to $|\mathbf{\Psi}|$ Q-values, one per pilot action. It relies on identical \ac{Eval-Net} and \ac{Target-Net} architectures, with input and output dimensions given by the numbers of communication beams and pilots, $\mathrm{n}_{\text{in}}$ and $\mathrm{n}_{\text{out}}$, respectively. Training uses \ac{MSE} loss, experience replay, $\epsilon$-greedy exploration with exponential decay $\lambda_\epsilon=0.992$, soft target updates with $\tau=0.005$, gradient clipping with norm 0.5 and Adam optimizer with learning rate $\alpha=10^{-4}$. Table~\ref{tab:hparams} summarizes the adopted hyperparameters. With these settings, the online cost derived in Section~\ref{sec:complexity} amounts to about $1.4\times10^5$ and $1.6\times10^5$ floating-point operations per decision for $(\mathrm{N_c},\mathrm{N_p})=(8,14)$ and $(16,30)$, respectively. A reduced discount factor $\gamma=0.94$ is used to favor shorter paths under negative step rewards, consistent with findings in~\cite{amit2020discount} which suggest that lower discount factors can improve generalization through implicit regularization. At test time, the selected action is the index of the maximum output Q-value which corresponds to the optimal configuration to be used. 
\begin{table}[!b]
    \centering
    \caption{Hyper-parameters Descriptions.}
    \renewcommand{\arraystretch}{1.1}
    \resizebox{0.95\columnwidth}{!}{
    \begin{tabular}{|c|c|}
    \hline
    \textbf{Component} & \textbf{Value / Setting} \\
    \hline
    $\mathrm{N}_{\min}$ & 200 \\
    \hline
    Number of layers $\mathrm{h}$ & 2 \\
    \hline
    Number of neurons per layer & 256 \\
    \hline
    Activation, norm & ReLU, LayerNorm\\
    \hline
    Episodes & $6{,}000$ \\
    \hline
    Channels per episode & 5 \\
    \hline
    Time steps per channel & 64 \\
    \hline
    Maximum path length $\mathrm{L^{*}}$ per time step & 20 \\
    \hline
    Replay buffer memory size & 120000 \\
    \hline
    Batch size & 1024 \\
    \hline
    Exploration & $\epsilon$-greedy\\
    \hline
    Initial exploration rate $\epsilon$ & 1 \\
    \hline
    Minimum exploration rate $\epsilon_{\text{min}}$ & 0.01 \\
    \hline
    Exploration decay rate $\lambda_{\epsilon}$  & 0.992 \\
    \hline
    Target network soft update & $\tau=0.005$ \\
    \hline
    Gradient clipping & Global-norm $0.5$ \\
    \hline
    Discount factor $\gamma$ & $0.94$ \\
    \hline
    Loss function & MSE \\
    \hline
    Optimizer / Learning Rate & Adam, $\alpha=1\times10^{-4}$\\
    \hline
    Experiment seed & $42$\\
    \hline
    $\delta$ value & $0.01$\\
    \hline
    Precision $1 - \delta$ & $0.99$\\
    \hline
    Hardware & CPU model \\
    \hline
    \end{tabular}}
\label{tab:hparams}
\end{table}
\subsection{Comparison with benchmarks}
We compare to the following benchmarks:




\begin{itemize}
    \item \textbf{Exhaustive Search:} Tests all $\mathrm{N_c}$ codewords in the communication codebook and selects the one that maximizes the objective in~(\ref{eq:optim_pb}).
    \item \textbf{Hierarchical Search:} A dichotomic procedure using $2\log_2(\mathrm{N_c})$ codewords in our setup. It starts with wide beams and progressively narrows the search to identify the optimal narrow beam. 
    \item \textbf{Q-Learning Acquisition:} Learns an optimal pilot-selection policy with the Q-table.
\end{itemize}
The experiments are repeated over many channel realizations and the results are averaged to ensure statistical reliability.
\begin{figure}
    \centering
    \includegraphics[width = 0.5\textwidth]{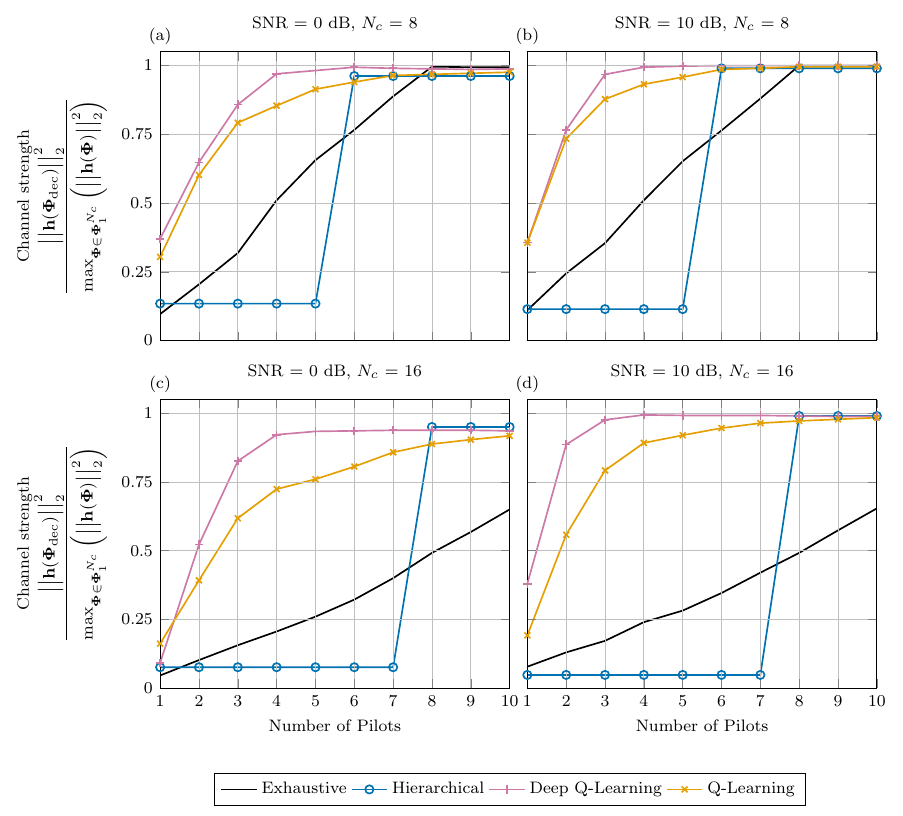}
    \caption{Normalized channel strength versus pilots for different methods, codebook sizes, and SNRs.}    
\label{fig:model_comparison_snr}
\end{figure}
\subsection{Convergence Behavior and Efficiency Analysis}
Figure~\ref{fig:model_comparison_snr} shows the normalized channel strength versus the number of pilots for the proposed \ac{DQL} and benchmark methods, for $\mathrm{N_c}\!\in\!\{8,16\}$ and \ac{SNR}$\in\{0,10\}$ dB, with results averaged over 500 channel realizations per \ac{SNR}. 
In all cases, \ac{DQL} converges faster and achieves higher performance than the baselines. At 0 dB with $\mathrm{N_c}=8$, it exceeds 97\% of the optimal gain after only 4 pilots, about half as many as competing methods need, and at 10 dB it needs just 3 pilots. Increasing the codebook size to $\mathrm{N_c}=16$ preserves this behavior (92\% after 4 pilots at 0 dB), showing good scalability to larger action spaces. Overall, \ac{DQL} learns more efficient pilot-selection policies, reducing pilot overhead across different \ac{SNR} and codebook sizes.

\begin{figure}[h]
    \centering
    \includegraphics[width= 0.3\textwidth]{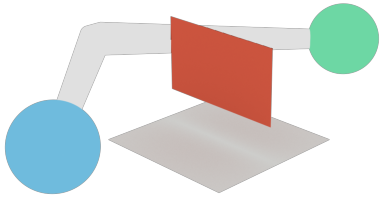}
    \caption{3D Sionna simulation setup}
    \label{fig:scenario_sionna}
\end{figure}

\emph{Discussion and future work:}
This work provides an initial validation of the proposed \ac{DQL} framework under a simplified single-path setting. As a next step, we are extending this study to larger and more structured pilot and communication codebooks, varying \ac{RIS} sizes, and more realistic Sionna-generated datasets based on multipath channel models~\cite{hoydis2022sionna}, in order to further assess generalization, scalability, and practical relevance. Fig.~\ref{fig:scenario_sionna} illustrates the considered Sionna setup, where the BS is shown in blue, the UE in green, and the RIS interaction is represented through the reflected paths. Preliminary results are promising and will be reported in a forthcoming publication. The source code and datasets for this work will also be released as open source at \url{github.com/OumaimaBounhar}.
\label{Conclusion}
\section{Conclusion}
We proposed a Deep Reinforcement Learning (\ac{DRL}) framework to optimize Reconfigurable Intelligent Surfaces (\ac{RIS}) configuration in a blind \ac{mmWave} Multiple-Input Multiple-Output (MIMO) communication, aiming to reduce the number of required pilots while preserving near-optimal beam selection. The proposed method consistently outperforms Q-Learning (\ac{QL}) and conventional search strategies, showing faster convergence and stronger robustness across both low- and moderate-\ac{SNR} regimes, and for both small and larger codebooks. In particular, the model reaches more than \textbf{97\%} (resp. \textbf{92\%}) of the optimal normalized channel strength after only 4 pilots at 0 dB for 8 (resp. 16) communication beams, and approaches near-optimal performance with just 3 pilots at 10 dB. These results confirm that the proposed approach remains effective as the codebook size grows, substantially reducing beam training overhead while maintaining strong beam-selection performance.

\bibliographystyle{ieeetr}
\bibliography{paper_bibtex.bib}

\end{document}